\documentclass[conference, a4paper]{IEEEtran}

\usepackage{algorithm,algorithmic}
\usepackage{amsmath,amssymb,mathrsfs}
\usepackage{array}
\usepackage{balance}
\usepackage{booktabs}
\usepackage{cite}
\usepackage{epsfig}
\usepackage{easyReview}
\usepackage{fancyhdr}
\usepackage{float}
\usepackage[nomain,acronym,automake,toc,nopostdot]{glossaries}
\usepackage{graphicx,color}
\usepackage{mathtools}
\usepackage{multirow}
\usepackage{psfrag}
\usepackage{stfloats}
\usepackage{subfigure}
\usepackage[left=1.44cm,right=1.44cm,top=2.88cm]{geometry}

\newtheorem{rem}{Remark}

\newtheorem{thm}{Theorem}

\makeglossaries
\newacronym{1d}{1D}{one-dimensional}
\newacronym{2d}{2D}{two-dimensional }
\newacronym{3d}{3D}{three-dimensional}
\newacronym{3gpp}{3GPP}{3rd Generation Partnership Project}
\newacronym{5g}{5G}{fifth-generation}
\newacronym{6g}{6G}{sixth-generation}

\newacronym{ap}{AP}{access point}
\newacronym{amp}{AMP}{approximate message passing}
\newacronym{ask}{ASK}{amplitude shift keying}
\newacronym{awgn}{AWGN}{additive white Gaussian noise}

\newacronym{ber}{BER}{bit error rate}
\newacronym{bfgs}{BFGS}{Broyden–Fletcher–Goldfarb–Shanno}
\newacronym{bler}{BLER}{bit error rate}
\newacronym{bp}{BP}{belief propagation}
\newacronym{bpsk}{BPSK}{binary phase shift keying}
\newacronym{bs}{BS}{base station}

\newacronym{cbsm}{CBSM}{correlation-based stochastic model}
\newacronym{csirs}{CSI-RS}{channel state information reference signal}
\newacronym{cdf}{CDF}{cumulative distribution function}
\newacronym{cdma}{CDMA}{code division multiple access}
\newacronym{cg}{CG}{conjugate gradient descent}
\newacronym{clt}{CLT}{central limit theorem}
\newacronym{csi}{CSI}{Channel state information }
\newacronym{cvp}{CVP}{closest vector problem}

\newacronym{dof}{DoF}{degrees of freedom}

\newacronym{edw}{EDW}{exponentially decaying window}
\newacronym{elaa}{ELAA}{extremely large aperture array}
\newacronym{etsi}{ETSI}{European Telecommunications Standards Institute}

\newacronym{ff}{FF}{far-field}
\newacronym{fcsd}{FCSD}{fixed-complexity sphere decoder}
\newacronym{fec}{FEC}{forward error correction}
\newacronym{fspl}{FSPL}{free space path loss}

\newacronym{gbsm}{GBSM}{geometry-based stochastic model}
\newacronym{gd}{GD}{gradient descent}
\newacronym{gmsk}{GMSK}{Gaussian minimum shift keying}
\newacronym{gs}{GS}{Gauss-Seidel}
\newacronym{gsm}{GSM}{global system for mobile communication}

\newacronym{iid}{i.i.d.}{independently and identical distributed}
\newacronym{ils}{ILS}{integer least-squares}
\newacronym{ind}{i.n.d.}{independently and non-identical distributed}
\newacronym{imt}{IMT}{International Mobile Telecommunications}
\newacronym{isac}{ISAC}{integrated sensing and communication}
\newacronym{isi}{ISI}{intersymbol interference}
\newacronym{itur}{ITU-R}{International Telecommunication Union Radiocommunication Sector}
\newacronym{iui}{IUI}{inter-user interference}

\newacronym{ji}{JI}{Jacobi iteration}
\newacronym{jsac}{JSAC}{joint sensing and communication}

\newacronym{las}{LAS}{likelihood ascent search}
\newacronym{lbfgs}{LBFGS}{limited-memory Broyden–Fletcher–Goldfarb–Shanno}
\newacronym{lll}{LLL}{Lenstra-Lenstra-Lov\'{a}sz}
\newacronym{llr}{LLR}{log-likelihood ratio}
\newacronym{los}{LoS}{line of sight}
\newacronym{lr}{LR}{lattice reduction}
\newacronym{lmmse}{LMMSE}{minimum mean square error}
\newacronym{lsd}{LSD}{list sphere decoder}
\newacronym{lte}{LTE}{long-term evolution}

\newacronym{map}{MAP}{maximum a posteriori}
\newacronym{mf}{MF}{matched filter}
\newacronym{mimo}{MIMO}{multiple-input multiple-output}
\newacronym{mld}{MLD}{maximum likelihood detection}
\newacronym{mmimo}{mMIMO}{massive multiple-input multiple-output}
\newacronym{mrc}{MRC}{maximum ratio combining}
\newacronym{mse}{MSE}{mean square error}
\newacronym{ms}{MS}{matrix-splitting}
\newacronym{mt}{MT}{mobile terminal}
\newacronym{nf}{NF}{near-field}
\newacronym{nlos}{NLoS}{non-LoS}

\newacronym{od}{OD}{orthogonality defect}

\newacronym{pdf}{PDF}{probability distribution function}
\newacronym{pda}{PDA}{probabilistic data association}
\newacronym{pep}{PEP}{pairwise error probability}
\newacronym{pl}{PL}{path-loss}
\newacronym{pmf}{PMF}{probability mass function}
\newacronym{pwm}{PWM}{plane-wave model}

\newacronym{qam}{QAM}{quadrature amplitude modulation}
\newacronym{qpsk}{QPSK}{quadrature phase shift keying}
\newacronym{qn}{QN}{quasi-Newton}
\newacronym{rts}{RTS}{reactive tabu search}
\newacronym{ri}{RI}{Richardson iteration}
\newacronym{ris}{RIS}{reconfigurable intelligent surface}
\newacronym{rss}{RSS}{received signal strength}
\newacronym{rzf}{RZF}{regularized-ZF}

\newacronym{sa}{SA}{Seysen's algorithm}
\newacronym{sd}{SD}{steepest descent}
\newacronym{sdr}{SDR}{semidefinite relaxation}
\newacronym{ser}{SER}{symbol error rate}
\newacronym{sf}{SF}{shadow fading}
\newacronym{sic}{SIC}{succesive interference cancellation}
\newacronym{sinr}{SINR}{signal-to-interference-plus-noise ratio}
\newacronym{siso}{SISO}{single input single output}
\newacronym{snr}{SNR}{signal-to-noise ratio}
\newacronym{sns}{SNS}{spatial non-stationarity}
\newacronym{sota}{SoTA}{state-of-the-art}
\newacronym{ssor}{SSOR}{symmetric successive over-relaxation}
\newacronym{svd}{SVD}{singular value decomposition }
\newacronym{swm}{SWM}{spherical-wave model}

\newacronym{tr}{TR}{Technical Report}
\newacronym{ts}{TS}{tabu search}

\newacronym{uca}{UCA}{uniform cylindrical array}
\newacronym{ue}{UE}{user equipment}
\newacronym{ula}{ULA}{uniform linear array}
\newacronym{ura}{URA}{uniform rectangular array}
\newacronym{uma}{UMa}{urban macro}
\newacronym{umi}{UMi}{urban micro}
\newacronym{upa}{UPA}{uniform planar array}

\newacronym{vblast}{V-BLAST}{vertical Bell Labs layered space-time}

\newacronym{xlmimo}{XL-MIMO}{extra-large multiple-input multiple-output}

\newacronym{zf}{ZF}{zero-forcing}

\definecolor{sblue}{RGB}{0,51,120}
\definecolor{sred}{RGB}{200,51,130}

\newcommand{\figref}[1]{Fig. \ref{#1}}

\renewcommand{\eqref}[1]{(\ref{#1})}

\ifCLASSINFOpdf
\else
\fi

\begin{document}
\title{Fast Iterative ELAA-MIMO Detection Exploiting Static Channel Components}
\author{Jiuyu Liu, Yi Ma, and Rahim Tafazolli\\
	{\small 5GIC and 6GIC, Institute for Communication Systems, University of Surrey, Guildford, UK, GU2 7XH}\\
	{\small Emails: (jiuyu.liu, y.ma, r.tafazolli)@surrey.ac.uk}}
\markboth{}%
{}

\maketitle

\begin{abstract}
\Gls{elaa} is a promising \gls{mimo} technique for next generation mobile networks.
In this paper, we propose two novel approaches to accelerate the convergence of  current iterative MIMO detectors in ELAA channels.
Our approaches exploit the static components of the ELAA channel, which include \gls{los} paths and deterministic \gls{nlos} components due to channel hardening effects.
This paper proposes novel convergence acceleration techniques for fast iterative ELAA-MIMO detection by leveraging the static channel component, including the LoS paths and deterministic \gls{nlos} components that arise due to channel hardening.
Specifically, these static channel components are utilized in two ways: as preconditioning matrices for general iterative algorithms, and as initialization for \gls{qn} methods.
Simulation results show that the proposed approaches converge significantly faster compared to current iterative MIMO detectors, especially under strong LoS conditions with high Rician K-factor.
Furthermore, QN methods with the proposed initialization matrix consistently achieve the best convergence performance while maintaining low complexity.
\end{abstract}
\glsresetall

\section{Introduction}\label{sec1}
\Gls{elaa} is anticipated to be one of the key technologies for the next-generation \gls{mimo} system to enable massive communication services \cite{ITUR2023}.
This necessitates highly scalable detectors capable of handling the massive data streams associated with large MIMO configurations.
The \gls{mld}, while optimal, is not scalable with increasing MIMO size since its computational complexity grows exponentially with the number of data streams  \cite{Albreem2019}.
Instead, linear \gls{mimo} detectors like \gls{zf} and \gls{lmmse} are employed to provide near-\gls{mld} performance with lower complexity.
However, they both introduce substantial processing latency, as the channel matrix inversion must be performed sequentially.
Iterative algorithms that bypass explicit matrix inversion can reduce complexity to a quadratic order, including \textit{1)} \gls{ms} based methods, \textit{2)} \gls{gd} methods, and \textit{3)} \gls{qn} methods \cite{Liu2023a}.
However, such iterative algorithms often suffer from slow convergence or even divergence in ELAA-MIMO systems due to channel ill-conditioning \cite{Liu2023a}.

In conventional MIMO systems, the far-field channel can be approximated as planar wave propagation.
As a result, the \gls{los} channel columns are extremely correlated, resulting in a rank-deficient channel matrix of rank one \cite{Cui2023}.
This makes far-field MIMO systems unsupportive of multiplexing techniques with strong LoS links.
However, ELAA-MIMO systems exhibit significantly different channel characteristics, especially regarding the direct LoS component.
Since users are located in the near-field of ELAA, the propagation no longer follows a planar wavefront but rather a spherical wavefront \cite{Ouyang2023}.
Consequently, ELAA channels can provide a higher or even full \gls{dof} to support spatial multiplexing techniques, even in LoS conditions.
Nevertheless, the correlation still exists between columns in ELAA channel matrix, which significantly decreases the convergence speed of current iterative detection algorithms.
The unique near-field propagation characteristics of ELAA channels necessitate the development of novel fast iterative detectors.

In this paper, we propose to exploit the static channel component to accelerate the convergence of current iterative MIMO detectors. 
The static channel component consists of two parts: the direct \gls{los} paths and the statistical properties of the \gls{nlos} paths.
We propose two approaches to exploit this static channel component for faster convergence:
\textit{1)} Use it as a preconditioning matrix to reconstruct a system of linear equations with a better-conditioned transfer function. 
This can be applied to all three types of iterative MIMO detectors, i.e., \gls{gd}, \gls{qn}, and \gls{ms}-based methods.
\textit{2)} Use it as the initial Hessian matrix approximation, specific to QN methods.
Our simulation results demonstrate that both the proposed two approaches can significantly speed up the convergence of current iterative detectors
Furthermore, it is noteworthy that using the static channel component as the initial Hessian matrix can provide even faster convergence with lower computational complexity compared to the preconditioning approach.

\section{System Model, Preliminaries and Problem Statement}\label{sec2}
\subsection{System Model}
Suppose there are $M$ service antennas and $N$ user antennas in the MIMO system. The uplink signal transmission can be expressed as follows
\begin{equation}\label{eqn01}
	\mathbf{r} = \mathbf{H} \mathbf{s} + \mathbf{v},
\end{equation}
where $\mathbf{r} \in \mathbb{C}^{M \times 1}$ denotes the received signal vector, $\mathbf{H} \in \mathbb{C}^{M \times N}$ the random channel matrix, $\mathbf{s} \in \mathbb{C}^{N \times 1}$ the transmitted signal vector,
$\mathbf{v} \sim \mathcal{CN}(0,\sigma_v^2\mathbf{I})$ the \gls{awgn}, and $\mathbf{I}$ is an identity matrix with compatible dimensions. 
MIMO channel in LoS condition is typically modeled as following a Rician distribution \cite{Liu2024a}
\begin{equation} \label{eqn02}
	\mathbf{H} = \sqrt{\dfrac{\kappa}{\kappa + 1}} \mathbf{H}_{\textsc{los}} + \sqrt{\dfrac{1}{\kappa + 1}} \mathbf{H}_{\textsc{nlos}},
\end{equation}
where $\kappa$ is the Rician K-factor; $\mathbf{H}_{\textsc{los}}$ and $\mathbf{H}_{\textsc{nlos}}$ denotes LoS and NLoS components, respectively.
In ELAA-MIMO systems,  the spherical wavefront propagation must be accounted for, as opposed to the planar approximation used in far-field. 
This indicates that the pathloss and phase need to be calculated based on the distance between each pair of service antenna and user antenna elements, as follows \cite{Liu2024a}
\begin{equation}\label{eqn03}
	h_{m,n}^{\textsc{los}} = \Bigg(\dfrac{\alpha}{d_{m,n}^{\beta}}\Bigg) \phi_{m,n}; \quad h_{m,n}^{\textsc{nlos}} = \Bigg(\dfrac{\alpha}{d_{m,n}^{\beta}}\Bigg) \omega_{m,n},
\end{equation}
where $\alpha$ denotes the pathloss coefficient, $\beta$ the pathloss exponent, $j$ the imaginary unit, and $d_{m,n}$ is the distance between the $m^{th}$ ELAA and $n^{th}$ user antennas;
$\phi_{m,n} = \exp(-j\frac{2\pi}{\lambda}d_{m,n})$ denotes the phase of direct LoS path;
$\omega_{m,n} \sim (0, 1)$ denotes the random variable following \gls{iid} Rayleigh fading.
The values of $\alpha$ and $\beta$ for different propagation scenarios are specified in the \gls{3gpp} technical documentation \cite{3gpp.38.901}.

\subsection{Linear MIMO Detectors and Iterative Algorithms} \label{sec02b}
The two most classic linear MIMO detectors are \gls{zf} and \gls{lmmse}.
Taking ZF as an example, its estimation can be expressed as follows
\begin{equation} \label{eqnzf}
	\mathbf{x} = (\mathbf{H}^{H} \mathbf{H})^{-1} \mathbf{H}^{H} \mathbf{r}.
\end{equation}
It is evident that the ZF detector requires inverting a Gram matrix, which can lead to substantial processing delays, making it challenging for real-time signal reception.

Let us define $\mathbf{A} \triangleq \mathbf{H}^{H}\mathbf{H}$ and $\mathbf{b} \triangleq \mathbf{H}^{H} \mathbf{r}$. 
Through simple linear transformations, \eqref{eqnzf} can be expressed as\footnote{By replacing $\mathbf{A} = \mathbf{H}^{H}\mathbf{H} + \frac{\sigma_{v}^{2}} {\sigma_{s}^{2}} \mathbf{I}$, the equation becomes \gls{lmmse} detector.
Due to the page limitation, we are unable to show such result in this paper.
The results is under going in our transaction version.}
\begin{equation}\label{eqn04}
\mathbf{A} \mathbf{x} = \mathbf{b}.
\end{equation}
Iterative algorithms can be employed to determine the solution $\mathbf{x}$ while bypassing the need for explicitly computing the matrix inverse of $\mathbf{A}$.
One classical iterative approach is the \gls{ri}, as follows \cite{Tu2020}
\begin{equation} \label{eqn05}
	\mathbf{x}_{t + 1} = \mathbf{x}_{t} - \mathbf{g}_{t},
\end{equation}
where $\mathbf{g}_{t} = \mathbf{A} \mathbf{x}_{t} - \mathbf{b}$ represents the update direction for RI at iteration $t$.
RI only requires matrix-vector multiplications, reducing the computational complexity to quadratic order.
Moreover, it inherently supports parallel implementations.
However, RI suffers from slow convergence when $\mathbf{A}$ is ill-conditioned.
This is because RI uses a fixed step size of $1$ and its update direction $\mathbf{g}_t$ may not provide sufficient descent, especially in ELAA-MIMO systems.
To accelerate convergence for ill-conditioned problems, three main categories of enhanced iterative algorithms have been developed.

\subsubsection{\Gls{ms}-Based Methods}
To provide better update directions and accelerate the convergence of \gls{ri}, MS-based methods were proposed. These methods split the matrix $\mathbf{A}$ as follows
\begin{equation}\label{eqn06}
	\mathbf{A} = \mathbf{D} + \mathbf{L} + \mathbf{L}^H,
\end{equation}
where $\mathbf{D}$ is the diagonal part of $\mathbf{A}$, and $\mathbf{L}$ is the strict lower triangular part of $\mathbf{A}$.
MS-based methods utilize the inverse of part of $\mathbf{A}$ to accelerate convergence, as follows
\begin{equation}\label{eqn07}
	\mathbf{x}_{t+1} = \mathbf{x}_{t} + \mathbf{M}^{-1} \mathbf{g}_{t},
\end{equation}
where $\mathbf{M}$ is called preconditioning matrix.
Different MS-based methods define $\mathbf{M}$ differently, such as $\mathbf{M}_\text{Jac} = \mathbf{D}$ for \gls{ji}, $\mathbf{M}_{\textsc{gs}} = \mathbf{D} + \mathbf{L}$ for \gls{gs} method \cite{Zhang2021}, and $\mathbf{M}_{\textsc{ssor}} = \mathbf{M}_{\textsc{gs}} \mathbf{D}^{-1}  \mathbf{M}_{\textsc{gs}}^H$ for \gls{ssor} \cite{Xie2016}.

\subsubsection{\Gls{gd} Methods}
To address the slow convergence caused by the fixed step size in RI, \gls{gd} methods adaptively tune the step size.
Taking \gls{sd} method as an example, its iterative process is given by \cite{Qin2016}
\begin{equation} \label{eqn01440507}
	\mathbf{x}_{t+1} = \mathbf{x}_{t} - \zeta_{t} \mathbf{g}_{t},
\end{equation}
where $\zeta_t$ is the step size at iteration $t$, given by
\begin{equation} \label{eqn02540507}
	\zeta_{t} = \dfrac{\mathbf{g}_{t}^{H} \mathbf{g}_{t}}{\mathbf{g}_{t}^{H} \mathbf{A} \mathbf{g}_{t}}.
\end{equation}
Clearly, \gls{ri} is the special case that $\zeta_{t} = 1$ for all the iterations.

\subsubsection{\Gls{qn} Methods}
\gls{qn} methods optimize both the update direction and step size for faster convergence. 
Their general iterative form is as follows \cite{Li2022a}
\begin{equation} \label{eqnqn}
	\mathbf{x}_{t+1} = \mathbf{x}_{t} - \gamma_{t} \mathbf{d}_{t},
\end{equation}
where the step size is given by \cite{Li2022a}
\begin{equation} \label{eqn03100507}
	\gamma_{t} = \dfrac{\mathbf{g}_{t}^{H} \mathbf{d}_{t}}{\mathbf{d}_{t}^{H} \mathbf{A} \mathbf{d}_{t}},
\end{equation}
where $\mathbf{d}_t = \mathbf{F}_t \mathbf{g}_t$ is the QN direction, with $\mathbf{F}_t$ approximating the inverse Hessian matrix.
Different QN methods update $\mathbf{F}_t$ differently.
For instance, $\mathbf{F}_{t}$ for \gls{lbfgs} is given by \cite{Liu2023a}
\begin{equation} \label{eqn09322704}
	\mathbf{F}_{t} = -\mathbf{F}_{0}+\dfrac{(\mathbf{x}_{t} - \mathbf{x}_{t-1})(\mathbf{g}_{t}-\mathbf{g}_{t-1})^{H}}{(\mathbf{x}_{t}-\mathbf{x}_{t-1})^{H}(\mathbf{g}_{t} - \mathbf{g}_{t-1})}\mathbf{F}_{0}.
\end{equation}
Usually, $\mathbf{F}_0$ is an identity matrix for MIMO detection \cite{Li2022}.

\subsection{Problem Statement}
The iterative methods discussed in Section \ref{sec02b} are generally applicable to any MIMO channel matrix. 
However, in ELAA-MIMO systems, the users are typically served under LoS conditions. 
In such scenarios, the high spatial correlation between the channel matrix columns may cause slow convergence for all three types of iterative algorithms \cite{Liu2024c}.
Moreover, although the GS and SSOR methods have quadratic complexity for inverting their preconditioning matrices, they do not support parallel computation. 
This can lead to substantial processing delays, which are particularly undesirable in real-time signal transmission where latency is a critical factor.

It is noteworthy that the direct LoS path remains unchanged as long as the relative positions of the ELAA array and the user are fixed \cite{Liu2021}. 
Therefore, we can assume that the inverse of the LoS channel component is available at the receiver-side, without introducing any additional signal processing delay. This key observation motivates the proposed approach in this paper.

\section{Fast Iterative Detection for ELAA-MIMO Systems Using Direct LoS Channel Component}
This section demonstrates how to leverage the static LoS channel component to accelerate the convergence of iterative MIMO detectors. 
We first split $\mathbf{A}$ into its LoS and NLoS components. 
Then, we propose two approaches - preconditioning and initialization - to incorporate the static channel component into iterations. 
Finally, a complexity analysis shows our methods are computationally efficient for practical implementation. 

\subsection{A Natural Way to Split $\mathbf{A}$}
In conventional MS-based methods, $\mathbf{A}$ is partitioned into its diagonal and non-diagonal parts.
However, this approach does not fully utilize the unique properties of ELAA channels. 
By plugging the Rician channel model into the definition of $\mathbf{A}$, we can express it as follows
\begin{IEEEeqnarray}{ll} \label{eqn14}
	\mathbf{A}\ &= \bigg(\dfrac{\kappa}{\kappa + 1}\bigg) \mathbf{H}_{\textsc{los}}^{H} \mathbf{H}_{\textsc{los}} + \bigg(\dfrac{1}{\kappa + 1}\bigg) \mathbf{H}_{\textsc{nlos}}^{H} \mathbf{H}_{\textsc{nlos}} \nonumber \\
	&+ \bigg(\dfrac{\sqrt{\kappa}}{\kappa + 1}\bigg) \mathbf{H}_{\textsc{los}}^{H} \mathbf{H}_{\textsc{nlos}} + \bigg(\dfrac{\sqrt{\kappa}}{\kappa + 1}\bigg) \mathbf{H}_{\textsc{nlos}}^{H} \mathbf{H}_{\textsc{los}}.
\end{IEEEeqnarray}
This shows $\mathbf{A}$ can be naturally split into the sum of four matrices in ELAA-MIMO systems. 
Similar to MS methods, we can use part of $\mathbf{A}$ as a preconditioning matrix to accelerate RI convergence. It is worth noting that not only the first term but also the second term can be utilized, since we have the following theorem.

\begin{thm} \label{thm01}
	Suppose that elements of $\mathbf{H}$ follow the Rician distribution given in \eqref{eqn02}, and given $N$, as $M$ approaches infinity, the Gram channel matrix $\mathbf{A}$ converges to
	\begin{equation} \label{eqn15}
		\lim\limits_{M \rightarrow \infty} \mathbf{A} = \Big(\dfrac{\kappa}{\kappa + 1}\Big) \mathbf{H}_{\textsc{los}}^{H} \mathbf{H}_{\textsc{los}} + \Big(\dfrac{1}{\kappa + 1}\Big) \mathbf{I}.
	\end{equation}
\end{thm}

\begin{IEEEproof}
	See \textsc{Appendix} \ref{app01}.
\end{IEEEproof}

\begin{rem}
	\textit{Theorem} \ref{thm01} shows that the Gram channel matrix $\mathbf{A}$ becomes a constant matrix in the limit of infinitely many serving antennas.
	Although this is not practically realizable, we can utilize this result to improve the convergence of current iterative algorithms.
	The justification is that ELAA-MIMO systems typically deploy a large number of antennas, ranging from hundreds to thousands of elements. 
	Therefore, let us define the following matrix
	\begin{equation} \label{eqn18}
		\mathbf{\Psi} = \bigg[\Big(\dfrac{\kappa}{\kappa + 1}\Big) 	\mathbf{H}_{\textsc{los}}^{H} \mathbf{H}_{\textsc{los}} + \Big(\dfrac{1}{\kappa + 1}\Big) \mathbf{I}\bigg]^{-1},
	\end{equation}
	which is called static channel component in ELAA-MIMO systems.
	Note that the Rician K-factor should be estimated in practice. 
	This is not the key contribution of this paper, so the robustness to inaccurate $\kappa$ will be addressed in future work.
\end{rem}

The matrix $\mathbf{\Psi}$ enables two approaches to accelerate iterative algorithms. 
Firstly, the preconditioning approach uses $\mathbf{\Psi}$ as a preconditioner to accelerate convergence. 
Secondly, the initialization approach uses $\mathbf{\Psi}$ as the initial Hessian matrix for QN methods. 
The next two subsections will present step-by-step details on the preconditioning and initialization methods.

\subsection{The Proposed Preconditioning Approach}
The proposed preconditioning approach can accelerate all three types of current iterative algorithms.

\subsubsection{Preconditioned RI (P-RI)}
Straightforwardly, $\mathbf{\Psi}$ can be used to accelerate RI as follows
\begin{equation} \label{eqn02040507}
	\mathbf{x}_{t+1} = \mathbf{x}_{t} - \mathbf{\Psi} \mathbf{g}_{t},
\end{equation}
which shares a similar form as \eqref{eqn07}. 
Our simulations in Section \ref{sec04} confirm the proposed method converges faster than GS and SSOR methods.

\subsubsection{Preconditioned GD Methods}
Some previous work combined GD and MS-based methods \cite{Qin2016}.
This motivates us to combine P-RI with current GD methods (e.g., SD methods).
Multiplying $\mathbf{\Psi}$ with both sides of \eqref{eqn04}, we have
\begin{equation} \label{eqn01360507}
	\mathbf{\Psi} \mathbf{A} \mathbf{x} = \mathbf{\Psi} \mathbf{b} \quad \Longrightarrow \quad \mathbf{\bar{A}} \mathbf{x} = \mathbf{\bar{b}},
\end{equation}
where $\mathbf{\bar{A}} \triangleq \mathbf{\Psi} \mathbf{A}$ and $\mathbf{\bar{b}} \triangleq \mathbf{\Psi}\mathbf{b}$.
With this linear function,  SD method in \eqref{eqn01440507} can also be used to determine $\mathbf{x}$ as follows
\begin{equation} \label{eqn01450507}
	\mathbf{x}_{t+1} = \mathbf{x}_{t} - \bar{\zeta}_{t} \mathbf{\bar{g}}_{t},
\end{equation}
where $\mathbf{\bar{g}}_{t} = \mathbf{\Psi} \mathbf{g}_{t}$  is the update direction, and the step size is as follows
\begin{equation} \label{eqn02590507}
	\bar{\zeta}_{t} = \dfrac{\mathbf{\bar{g}}_{t}^{H} \mathbf{\bar{g}}_{t}}{\mathbf{\bar{g}}_{t}^{H} \mathbf{\bar{A}} \mathbf{\bar{g}}_{t}}.
\end{equation}
This is referred to as preconditioned SD (P-SD) method.

\subsubsection{Preconditioned QN Methods}
\Gls{qn} methods can also be used to determine $\mathbf{x}$ according to \eqref{eqn01360507}.
Taking \gls{lbfgs} as an example, its iterative process can be expressed as follows
\begin{equation} \label{eqn01540507}
	\mathbf{x}_{t+1} = \mathbf{x}_{t} - \bar{\gamma}_{t} \mathbf{\bar{d}}_{t}.
\end{equation}
Accordingly, $\bar{\gamma}_{t}$ can be expressed as follows
\begin{equation} \label{eqn03170507}
	\bar{\gamma}_{t} = \dfrac{\mathbf{\bar{g}}_{t}^{H} \mathbf{\bar{d}}_{t}}{\mathbf{\bar{d}}_{t}^{H} \mathbf{\bar{A}} \mathbf{\bar{d}}_{t}}.
\end{equation}
The update direction is $\mathbf{\bar{d}}_{t} = \mathbf{F}_{t} \mathbf{\bar{g}}_{t}$, where $\mathbf{F}_{t}$ for \gls{lbfgs} updates is given as follows
\begin{equation} \label{eqn02010507}
	\mathbf{F}_{t} = -\mathbf{F}_{0}+\dfrac{(\mathbf{x}_{t} - \mathbf{x}_{t-1})(\mathbf{\bar{g}}_{t}-\mathbf{\bar{g}}_{t-1})^{H}}{(\mathbf{x}_{t}-\mathbf{x}_{t-1})^{H}(\mathbf{\bar{g}}_{t} - \mathbf{\bar{g}}_{t-1})}\mathbf{F}_{0}.
\end{equation}
This preconditioned QN method using \gls{lbfgs} updates is referred to as the P-LBFGS method.

\subsection{The Proposed Initialization Approach}
The preconditioning methods propose above can accelerate the convergence but with extra computational cost.
QN methods iteratively approximate the inverse Hessian matrix using gradient/solution updates. 
Their convergence is faster if initialized closer to $\mathbf{A}^{-1}$ than the conventional identity initialization.
We propose to set $\mathbf{F}_0 = \mathbf{\Psi}$, which is the static channel component, to accelerate the convergence of QN methods. 
For LBFGS, this leads to the following update
\begin{equation} \label{eqn02140507}
	\mathbf{F}_{t} = -\mathbf{\Psi} + \dfrac{(\mathbf{x}_{t} - \mathbf{x}_{t-1})(\mathbf{g}_{t}-\mathbf{g}_{t-1})^{H}}{(\mathbf{x}_{t}-\mathbf{x}_{t-1})^{H}(\mathbf{g}_{t} - \mathbf{g}_{t-1})} \mathbf{\Psi}.
\end{equation}
This approach shares the same formulation as the original LBFGS method, with the only difference being the initial inverse Hessian approximation $\mathbf{F}_0$.
This method is refer to as I-LBFGS method for convenience in demonstrating complexity and performance.
 
\subsection{Complexity Analysis}
The computational complexity analysis will be presented in this section.
The preprocessing of calculating $\mathbf{A}$ and $\mathbf{b}$ is ignored since it is present in all methods that support parallel computation.
Additionally, $\mathbf{\Psi}$ is a constant matrix and does not vary with the channel state information.
Moreover, the complexity of addition and subtraction operations is considered negligible in this analysis.

Our complexity analysis starts with the simple \gls{ri}, where the only complexity arises from calculating $\mathbf{A}\mathbf{x}_t$ with $N^2$ complexity.
For the proposed method in \eqref{eqn02040507}, it requires calculating $\mathbf{\Psi} \mathbf{g}_t$, which has an additional complexity of $N^2$.
The preconditioning matrix for JI is a diagonal matrix, so that it requires an additional complexity of $2N$ to calculate $\mathbf{D}^{-1}$ and $\mathbf{D}^{-1} \mathbf{g}_{t}$. 
Both GS and SSOR methods require the inverse of $\mathbf{M}_\textsc{gs}$, which has a serial computational complexity of $N^2$.
Furthermore, the complexities of calculating $\mathbf{M}^{-1} \mathbf{g}_t$ for GS and SSOR methods are $N^2$ and $2N^2 + N$, respectively.
For the SD method, it also requires calculating $\mathbf{g}_t$ with a complexity of $N^2$. 
Moreover, the calculation of $\zeta_{t}$ in \eqref{eqn02540507} requires an additional complexity of $N^2 + 2N$. 
Therefore, the overall complexity of the SD method is $2N^2 + 2N$.
For the proposed preconditioning methods combined with the SD method, calculating the new update direction $\mathbf{\bar{g}}_t$ requires an additional complexity of $N^2$.
Also, calculating the step size $\bar{\zeta}_t$ requires $2N^2 + 2N$.
Therefore, the complexity of the proposed SD method with preconditioning is $4N^2 + 2N$ per iteration.

For LBFGS method, it also requires calculating $\mathbf{g}_t$ with a complexity of $N^2$.
Then, calculating $\mathbf{F}_{t} \mathbf{g}_t$ requires a complexity of $N^2 + 2N$.
Moreover, calculating the step size in \eqref{eqn03100507} requires an additional complexity of $N^2 + 2N$.
Therefore, the overall complexity of the LBFGS method is $3N^2 + 4N$.
For the preconditioning approach with the LBFGS method, calculating $\mathbf{\bar{g}}_t$ requires an additional $N^2$ complexity, and calculating the new step size in \eqref{eqn03170507} requires an additional $\mathcal{O}(N^2)$ complexity compared to \eqref{eqn03100507}.
Therefore, the overall complexity of the proposed preconditioning approach for the LBFGS method is $5N^2 + 4N$.
Finally, the complexity of the proposed LBFGS method using $\mathbf{\Psi}$ as the initial Hessian approximation has the same complexity as the original LBFGS method, which is $3N^2 + 4N$.

The summary of all these iterative algorithms is presented in \textsc{TABLE} \ref{tab01}.
It can be observed that all the proposed methods have quadratic complexities, maintaining their computational efficiency for practical implementation.

\begin{table}[]
\caption{Complexity and Performance for Different Algorithms}
\label{tab01}
\centering
\renewcommand{\arraystretch}{0.7}
\resizebox{0.42\textwidth}{!}{
\begin{tabular}{@{}ccc@{}}
	\toprule
	\footnotesize
	Algorithms & Serial Complexity & Complexity Per Iteration \\ \midrule
	RI         & $0$                       & $N^2$                    \\
	JI         & $0$                       & $N^2 + 2N$               \\
	GS         & $\mathcal{O}(N^2)$        & $2N^2$                   \\
	SSOR       & $\mathcal{O}(N^2)$        & $3N^2 + N$               \\
	SD         & $0$                       & $2N^2 + 2N$              \\
	LBFGS      & $0$                       & $3N^2 + 4N$              \\
	P-RI       & $0$                       & $2N^2$                   \\
	P-SD       & $0$                       & $4N^2 + 2N$              \\
	P-LBFGS    & $0$                       & $5N^2 + 4N$              \\
	I-LBFGS    & $0$                       & $3N^2 + 4N$              \\ \bottomrule
\end{tabular}
}
\vspace{-1em}
\end{table}

\section{Simulation Results} \label{sec04}
This section aims to validate that the proposed preconditioning and initialization approaches can accelerate the convergence of current iterative MIMO detectors for ELAA channels.

\subsection{Experiment Setup}
In the computer simulations, we consider an ELAA-MIMO system where the service antennas are arranged in a large ULA with half-wavelength spacing at a central frequency of $3.5$ $\mathrm{GHz}$.
The modulation order is set to be $16$ QAM.
The ELAA is assumed to have $256$ or $512$ antennas. 
There are $8$ user equipments also deployed linearly with equal spacing, parallel to the large ULA, with a maximum distance of ten meters between the farthest users. 
Each user equipment is assumed to have four antennas with half-wavelength spacing.
The perpendicular distance between the users and the ELAA is set to thirty meters, clearly placing the users in the near-field of the ULA. 
The ELAA channels are assumed to follow Rician fading and are generated according to \eqref{eqn02}. 
To simulate scenarios with strong and weak LoS paths, the Rician K-factor is set to $\kappa = 8$ and $\kappa = 4$, respectively.

\subsection{Experiment  Results}
\begin{figure*}[t]
	\centering
	\subfigure[Strong LoS condition, $\kappa = 8$]{
	\begin{minipage}[t]{0.49\textwidth}	
			\label{fig01a}
			\centering
			\includegraphics[width=6.5cm]{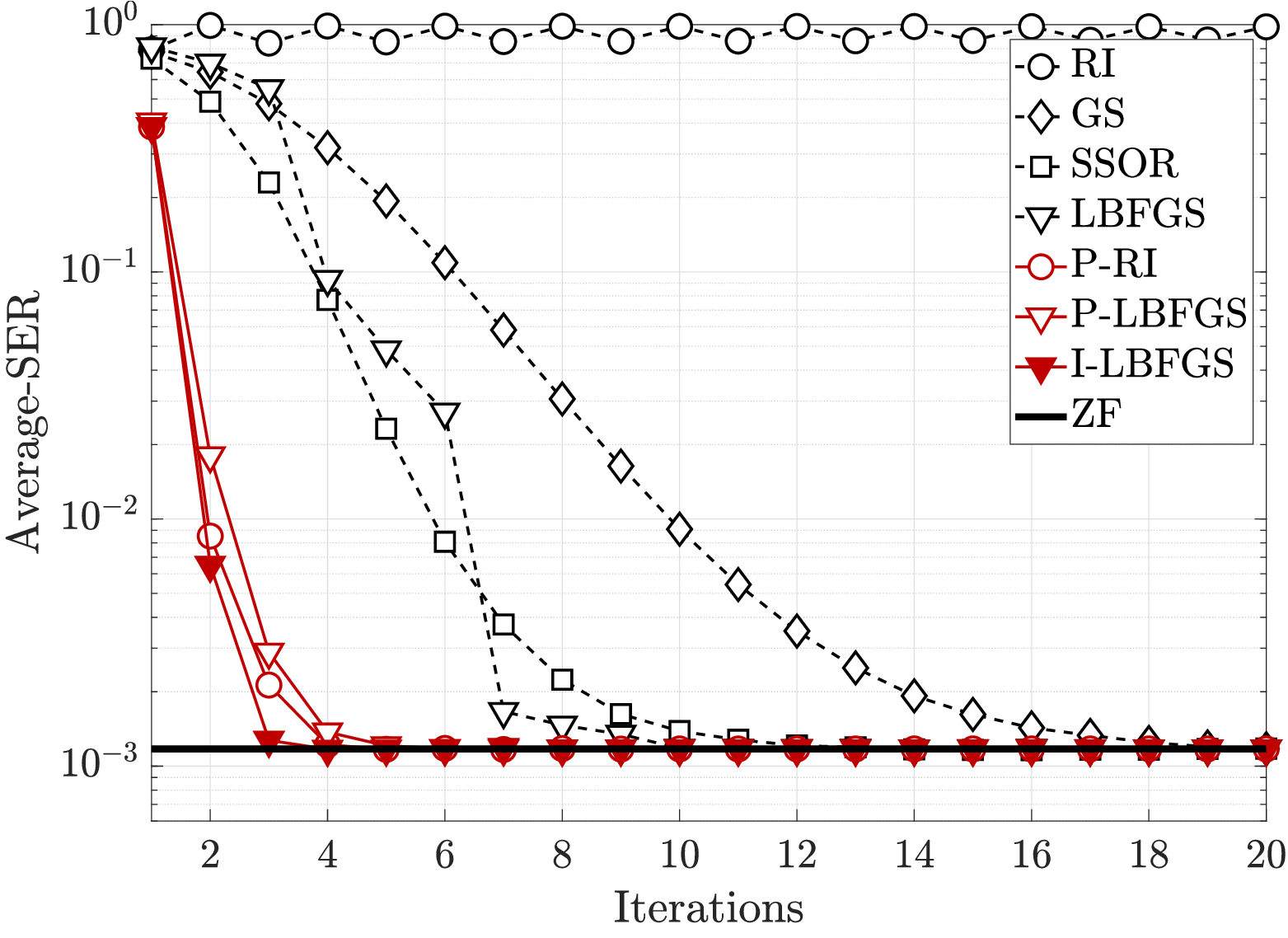}
			\vspace{0.5em}
	\end{minipage}}
	\subfigure[Weak LoS condition, $\kappa = 4$]{
		\begin{minipage}[t]{0.49\textwidth}
			\label{fig01b}
			\centering
			\includegraphics[width=6.5cm]{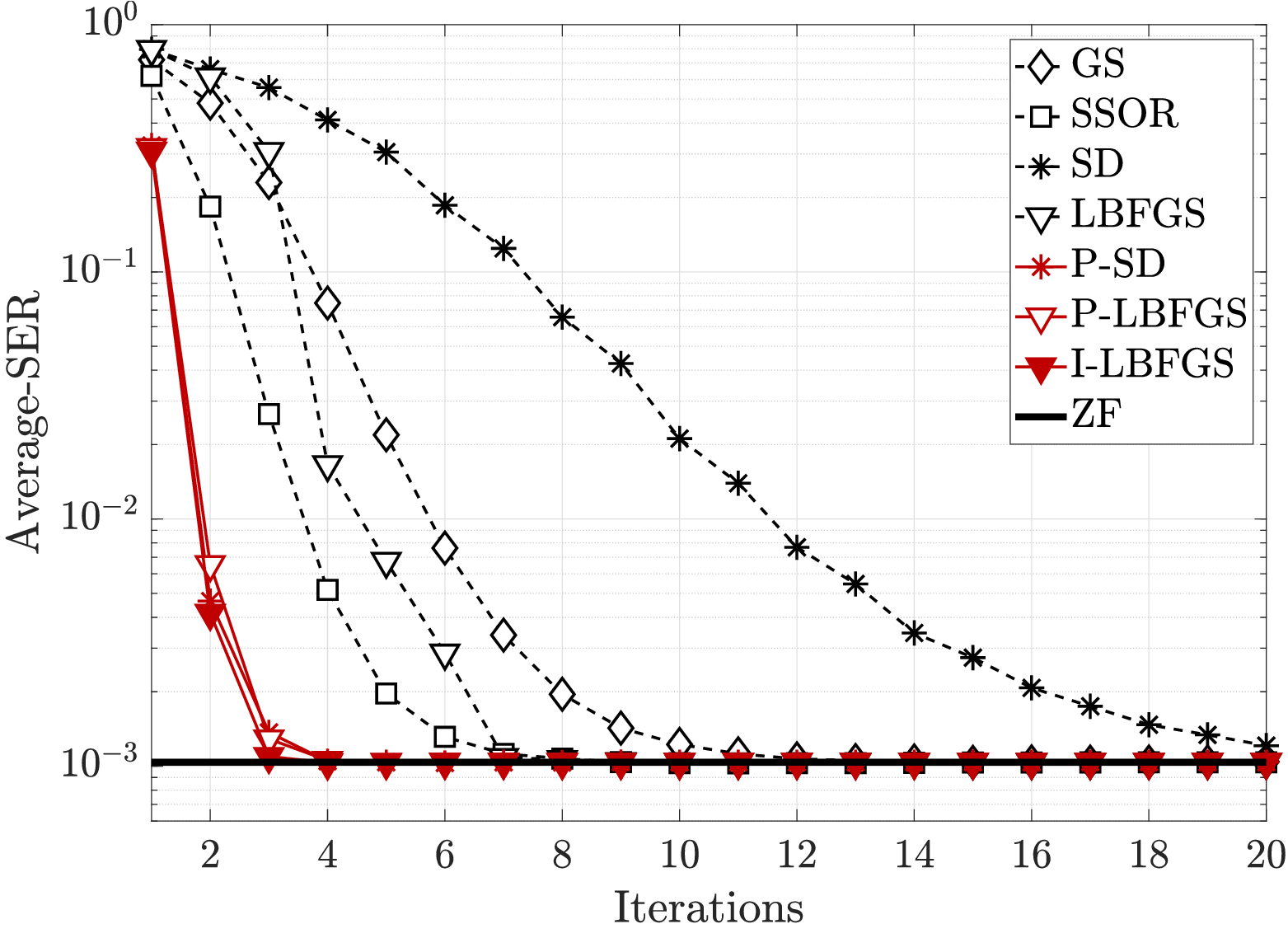}
			\vspace{0.5em}
	\end{minipage}}
	\caption{\label{fig01} Convergence behavior of different iterative MIMO detectors for ELAA systems. There are $512$ service antennas and $32$ user antennas in the system.}
	\vspace{-1em}
\end{figure*} 

\begin{figure}[t]
	\centering
	\begin{minipage}[t]{0.49\textwidth}	
			\centering
			\includegraphics[width=6.5cm]{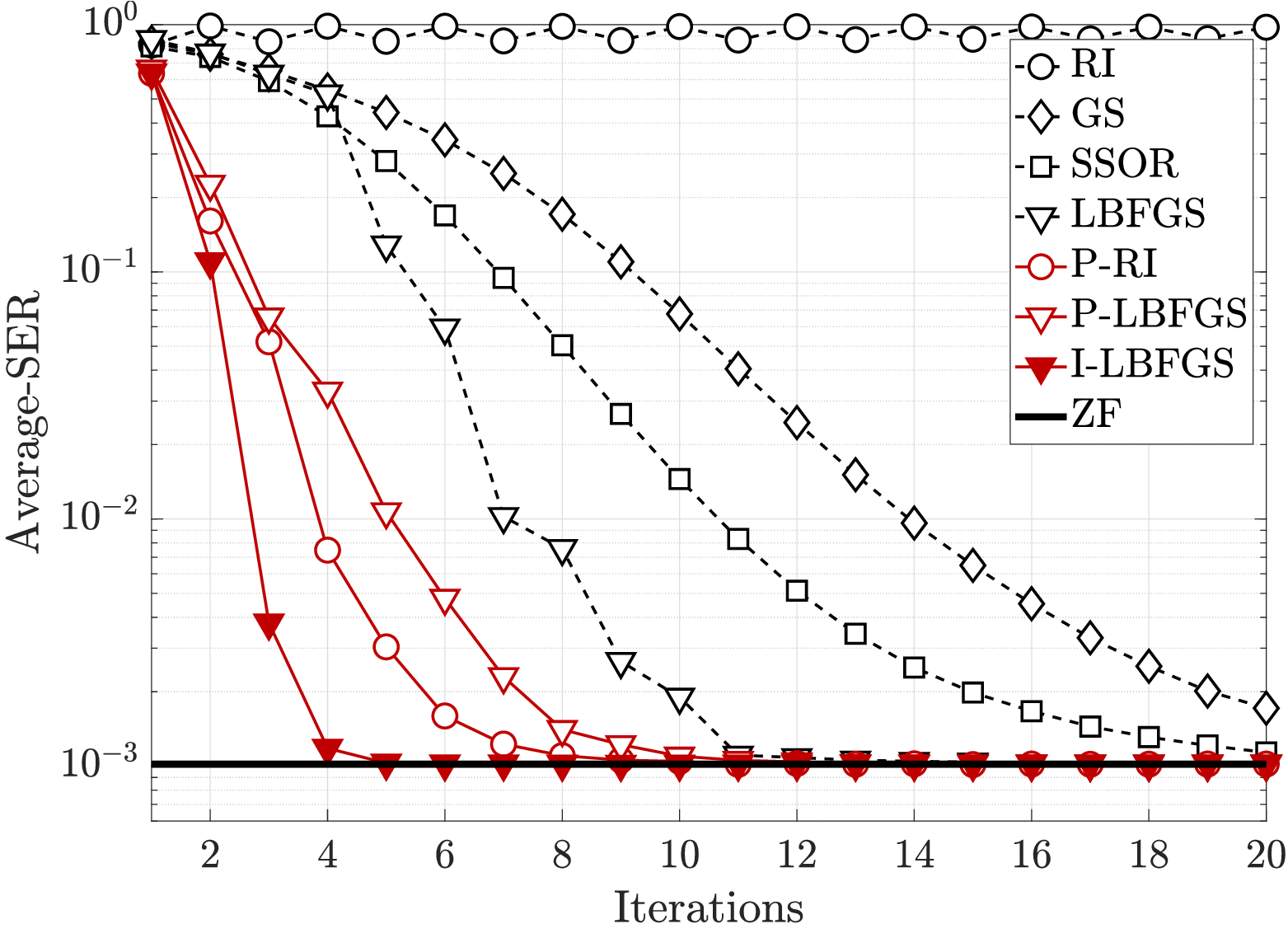}
	\end{minipage}
	\caption{\label{fig02} Convergence comparison in strong LoS condition ($\kappa = 8$) with $256$ service antennas and $32$ user antennas in the ELAA-MIMO system.}
	\vspace{-1em}
\end{figure}  

\figref{fig01} shows the convergence behavior of various iterative MIMO detection algorithms, including the proposed methods (P-RI, P-LBFGS, and I-LBFGS), for different Rician K-factor.
\figref{fig01a} corresponds to a Rician K-factor of $8$, which represents a strong \gls{los} channel condition. 
The proposed methods, P-RI, P-LBFGS, and I-LBFGS, exhibit significantly faster convergence compared to the conventional algorithms such as RI, GS, SSOR, and LBFGS.
This improvement in convergence speed can be attributed to the effective utilization of the direct LoS channel information in the proposed techniques.
\figref{fig01b} depicts the convergence performance for a Rician K-factor of $4$, which represents a weak LoS channel condition.
While the convergence speeds of current algorithms are faster compared to the strong LoS scenario in \figref{fig01a}, the proposed methods still demonstrate noticeable convergence acceleration over their conventional counterparts.
The proposed I-LBFGS method, which initializes the LBFGS algorithm with the direct LoS channel matrix, consistently achieves the fastest convergence among all iterative techniques across different K-factor values.
These results highlight the effectiveness of the proposed approaches in leveraging the static LoS channel information to improve the convergence rates of iterative MIMO detectors, especially in LoS-dominant ELAA-MIMO systems where the channel correlation effects are most prominent.
\figref{fig02} shows results for a smaller number of service antennas ($M=256$) under strong LoS conditions.
The proposed methods can still offer faster convergence in this scenario, further demonstrating their applicability across different ELAA system configurations.

\section{Conclusion}
This paper proposed two approaches that exploit the static channel component to accelerate convergence of iterative MIMO detectors for ELAA systems. 
The preconditioning approach uses the static component as a preconditioner to accelerate all the current iterative methods \gls{gd}, \gls{qn}, and \gls{ms}-based methods.
The initialization approach initializes the inverse Hessian approximation to be the static component, providing a better starting point compared to conventional identity initialization.
Simulation results were performed for an ELAA-MIMO system with $512$ antennas serving $8$ multi-antenna users in the near-field region. 
Results show that under strong LoS conditions, the proposed P-RI, P-SD and P-LBFGS methods converge substantially faster than conventional iterative detectors. 
Even in weak LoS scenarios, the proposed methods maintain convergence rate advantages.
This makes them promising techniques for low-complexity, low-latency signal detection in real-time ELAA-MIMO communications. 
Potential future work includes studying the robustness to imperfect LoS information and Rician K-factor, as well as extensions to LMMSE detector and other channel models.

\appendices
\section{Proof of \textit{Theorem} \ref{thm01}} \label{app01}
To prove \textit{Theorem \ref{thm01}}, we will use the law of large numbers and the properties of the Rician fading channel model.
Given the expression of $\mathbf{A}$ in \eqref{eqn14}, as $M \rightarrow \infty$, we analyze the convergence of each term in the expression as follows
The first term is deterministic and does not depend on $M$, so it remains unchanged as $M \rightarrow \infty$.
For the second term, we can apply the law of large numbers. 
As $M \rightarrow \infty$, the off-diagonal elements of this matrix converge to their expected value, which is zero due to the \gls{iid} nature of the NLoS components. 
Therefore, we have the following \cite{Ngo2014}
\begin{equation}
		\lim_{M \rightarrow \infty}  \mathbf{H}_{\textsc{nlos}}^H \mathbf{H}_{\textsc{nlos}} =  \mathbf{I}.
\end{equation}
For the cross terms (i.e., the third and forth terms), as $M \rightarrow \infty$, the elements of these matrices converge to their expected value, which is zero due to the independence between the LoS and NLoS components.
Combining these results, \eqref{eqn15} in \textit{Theorem} \ref{thm01} is therefore obtained.

\section*{Acknowledgement}
This work was partially funded by the 5G and 6G Innovation Centre, University of Surrey, and partially by the UK Department for Science, Innovation and Technology under the Future Open Networks Research Challenge project TUDOR (Towards Ubiquitous 3D Open Resilient Network).

\ifCLASSOPTIONcaptionsoff
\newpage
\fi

\bibliographystyle{IEEEtran}
\bibliography{../IEEEabrv,../mMIMO}
\end{document}